\documentstyle[aps,jpc,preprint]{revtex}

\begin{document}


\title{Hydrolysis of Ferric Ion in Water 
and Conformational Equilibrium}




\author{Richard L. Martin, P. Jeffrey Hay, and Lawrence R. Pratt }
\address{ Theoretical Division, Los Alamos National Laboratory, Los
Alamos, New Mexico 87545 USA }

\author{LA-UR-97-3489}

\date{\today}

\maketitle

\def\rtrade{{\sixrm $\bigcirc\!\!\!\!\!$R} } \def\ang{\AA$\;$}

\begin{abstract}

Reported here are results of theoretical calculations on
Fe(H$_2$O)$_6$$^{3+}$, Fe(H$_2$O)$_5$(OH)$^{2+}$, three isomers of
Fe(H$_2$O)$_4$(OH)$_2$$^{+}$, and Fe(H$_2$O)$_3$(OH)$_2$$^{+}$ to
investigate the molecular mechanisms of hydrolysis of ferric ion in
water.  The combination of density functional electronic structure
techniques and a dielectric continuum model for electrostatic
solvation applied to the Fe(H$_2$O)$_6$$^{3+}$ complex yields an
estimate of -1020~kcal/mol (experimental values -1037~kcal/mol to
-1019~kcal/mol) for the absolute free energy of the aqueous ferric
ion.  The free energy change for the first hydrolysis reaction is
predicted reasonably (2 kcal/mol predicted compared to 3 kcal/mol
experimental). For the second hydrolysis reaction, we found an
unexpected low energy isomer of Fe(H$_2$O)$_4$(OH)$_2$$^{+}$ with five
ligands in the inner sphere and one water outside. The hexa-coordinate
{\it cis\/} and {\it trans\/} isomers are, respectively, slightly
lower and higher in energy.  Calculations on the penta-coordinate
species Fe(H$_2$O)$_3$(OH)$_2$$^{+}$ suggest that extrusion of the
outer sphere water is nearly thermoneutral. The reaction free energy
for the second hydrolysis is predicted in the range 16-18~kcal/mol,
higher than the experimental value of 5~kcal/mol.  In view of the
facts that the theoretical predictions are higher than experimental
values, and that novel structures were encountered among products of
the second hydrolysis, we argue that conformational entropy is an
important omission in this theoretical treatment of net reaction free
energies. A fuller cataloging of low energy hydrolysis products and
direct calculations of partition functions of the isolated complexes
should help in modeling equilibrium speciation in groundwaters.

\end{abstract}
\pagebreak



\section{Introduction}

A molecular theoretical account of the free energetics of the
reactions
\begin{eqnarray}
\label{eq:hydro} [\mbox{Fe(H$_2$O)$_n$(OH)$_{(m-1)}$]$^{4-m}$} +
\mbox{H$_2$O} \rightarrow [\mbox{Fe(H$_2$O)$_{n-1}$(OH)$_{m}$]$^{3-m}$}
+ \mbox{H$_3$O$^+$}
\end{eqnarray} 
in water can benefit from discrimination of the conformers that are
present under common conditions.  Entropic contributions to the
solution thermodynamics may reflect multiple configurations that
occur.  Thus information on the conformers present should assist in
accurately describing temperature variations of solvation properties.
In addition, theoretical study of the molecular structure of the
species participating these reactions should teach us about the
molecular mechanisms involved and provide a benchmark of current
theoretical tools for modeling speciation of metal ions in
groundwaters\cite{Rustad:95,Rustad:96a,Rustad:96b}.

The condition of ferrous and ferric ions in solution has long been of
specific interest to explanations of electron exchange processes
\cite{RMarcus,Friedman,Jafri,Newton}.  The hydrolysis product
Fe(H$_2$O)$_5$OH$^{2+}$ often contributes significantly to the rate of
ferric ion reduction in water because the specific rate constant for
ferric-ferrous electron exchange is about a thousand times larger for
the hydrolyzed compared to unhydrolyzed hexaaquoferric ion
\cite{Silverman}. The net standard free energy change (about
3~kcal/mol\cite{Flynn}) for the first hydrolysis reaction,
\begin{eqnarray}
\label{eq:hydro1} \mbox{Fe(H$_2$O)$_6$$^{3+}$} + \mbox{H$_2$O}
\rightarrow \mbox{Fe(H$_2$O)$_{5}$(OH)$^{2+}$} + \mbox{H$_3$O$^+$} , 
\end{eqnarray} 
is small compared to the size of the various contributions that must
be considered in theoretical modeling of these solution species.

In contrast to the great volume of work on electron transfer involving
such species, reactions of particular interest here transfer a proton
from a water ligated to a Fe$^{3+}$ ion to a free water molecule.  The
second hydrolysis,
\begin{eqnarray}
\label{eq:hydro2} \mbox{Fe(H$_2$O)$_5$OH$^{2+}$} + \mbox{H$_2$O}
\rightarrow \mbox{Fe(H$_2$O)$_{4}$(OH)$_2$$^+$} + \mbox{H$_3$O$^+$},
\end{eqnarray} 
raises more seriously the possibility of participation by several
isomers in the mechanism and thermodynamics of these reactions.

This work takes ferric ion hydrolysis as a demonstration system for
determination of the utility of current theoretical tools for modeling
of speciation of metal ions in aqueous solution.  At this stage we
have not considered multiple metal center aggregates.  We report below
encouraging agreement with experimental thermochemistries but with
unexpected results that must be addressed for further progress in
describing these chemical systems in molecular terms.

\section{Approach}

The physical idea of treating an inner solvation shell on a different
footing from the rest of the solvent has considerable
precedent\cite{Friedman:73}. Friedman and Krishnan referred to these
approaches as {\it hybrid models} and criticized them on the grounds
that individual contributions to the thermodynamic properties of final
interest could not be determined with sufficient accuracy to draw new
conclusions from the final results.  However, particularly for highly
charged and chemically complex ions, the hybrid approaches  seem
indispensable \cite{Marcos}.  Computational and conceptual progress in
the 25 years since that Friedman-Krishnan review has made these
approaches interesting once again. In what follows, we first elaborate
on the calculational techniques used. An appendix discusses some of the
statistical thermodynamic issues underlying these hybrid models.

\subsection{Electronic Structure Calculations}

Reliable estimation of free energies of dissociation in aqueous media
requires a similarly reliable determination of the process in the gas
phase. Given the well known difficulties associated with electron
correlation in transition metal complexes, and our intent to extend
this work to larger and more complex systems, we chose to explore this
problem with the B3LYP hybrid density functional theory
(DFT)\cite{b3lyp,G94}. This approximation is an effective compromise
between accuracy and computational expense, and has been shown to give
usefully accurate predictions of metal-ligand bond energies in a
number of molecules\cite{ricca:MCO,russo:MF}. In the present
application however, we were less interested in the metal-ligand bond
energy than in the energies for successive elimination of protons from
the H$_2$O ligands bound to Fe$^{3+}$.  This may be a somewhat less
severe test of the functional. The details of our approach and
estimates of the errors follow.

The geometry and vibrational frequencies of the metal complex were
determined using the 6-311+G basis set for the metal ion and the 6-31G*
basis set for the ligands\cite{G94}. The former is contracted from
Wachters'\cite{wachters} primitive Gaussian basis (including the
functions describing the 4s and 4p atomic orbitals) and augmented with
the diffuse d-function of Hay\cite{hay-dif-d}. The ligand basis set
contains a polarization function of d-character on the oxygen. At the
optimum geometry, a single calculation was done to determine the energy
in a more extended basis (6-31++G**) that includes polarization and
diffuse functions on both the oxygen and hydrogen centers. Atomic
charges were determined in this extended basis as well using the
ChelpG\cite{chelpg} capability (R$_{Fe}$=2.02 \ang) in the Gaussian94
package\cite{G94}. All Fe species are high-spin (d$^5$) treated in the
spin-unrestricted formalism.

One might argue, on the basis of the OH$^-$ character expected in the
product complexes, that the geometry optimization should also be done
with a diffuse basis set for the ligands.  This was found unnecessary
by explicit calculations on the partial reaction
\begin{eqnarray}
\mbox{Fe(H$_2$O)$_6$$^{3+}$} \rightarrow
\mbox{Fe(H$_2$O)$_5$(OH)$^{2+}$} + \mbox{H$^+$}. \label{partial}
\end{eqnarray} 
The endothermicity of this reaction computed from a single point
calculation with the 6-31+G* basis at its optimum geometry differs by
only 0.2~kcal/mole from the result obtained with the smaller basis.
More important are additional polarization and diffuse functions on
the hydrogens; for example, neglecting zero-point corrections, the
B3LYP endothermicities are 29.2~kcal/mole(6-31G*),
25.4~kcal/mole(6-31+G*), and 28.1~kcal/mole(6-31++G**).

A final consideration regarding basis set convergence is the
importance of polarization (f-functions) on the metal center.  This
was examined by augmenting the 6-311+G Fe basis with two f-functions
($\alpha$ = 0.25,0.75) split about a value ($\alpha$ = 0.4) optimized
for Fe(H$_2$O)$_6$$^{3+}$. In this larger basis an endothermicity of
27.9~kcal/mole results compared to the value 28.1~kcal/mole obtained
without the f-functions. The correction to the second hydrolysis is
only slightly larger, $\sim$ 0.5~kcal/mole. We thus conclude that the
B3LYP deprotonation energies are nearly converged with respect to
further improvements in the basis set. A conservative estimate might
be associated with the change observed between the 6-311+G/6-31+G*
results and the 6-311+G(2f)/6-31++G** results, or about 2-3~kcal/mole.

Convergence of the results with respect to basis set does not address
the accuracy expected with the B3LYP method. In so far as the water
molecules retain their identity when bound to the metal ion, we expect
the error in the hydrolysis reaction to be approximated by the error
observed for the analogous reaction for a single H$_2$O molecule. The
procedure outlined above was therefore used to determine the energies
and zero-point energies (ZPE) for a number of species associated with
the neutral and ionic dissociation channels of H$_2$O.  They are
reported in Table~\ref{T_0}, and can be used to determine the
properties and reaction energies reported in Table~\ref{T_1}. The
H$_2$O thermochemistries shown there are in excellent agreement with
experiment. For example, the error associated with the reaction
\begin{eqnarray} 
\mbox{H$_2$O} \rightarrow \mbox{H$^+$} + \mbox{OH$^-$}
\label{partial:h} 
\end{eqnarray} 
is underestimated by less than 2~kcal/mole. The errors in the other
dissociation channels are similar.

\subsection{Electrostatic Solvation Calculations}

Electrostatic interactions of hydrated ferric ions with the aqueous
environment are expected to be of first importance in this solution
chemistry. These hydrolysis reactions are written so that other
contributions to the net free energy, {\it e.g.\/} packing, might
balance reasonably.  Thus, continuum dielectric solvation modeling is
a reasonable approach to studying these solution processes; it is
physical, computationally feasible, and can provide a basis for more
molecular theory \cite{Pratt:94,Tawa:94,Tawa:95,Hummer:95a,Corcelli,%
Hummer:96,Tawa:96,Hummer:97a,Pratt:97a,Hummer:97b,Hummer:97c}. The
dielectric model produces an approximation to the interaction part of
the chemical potential, $\mu^\ast $ of the solute; in notation used
below $\mu^\ast = -RT\ln \langle \exp(-\Delta U) \rangle_0$ where for
this case $\Delta U$ is the solute-solvent electrostatic interaction
potential energy and the brackets indicate the average over the
thermal motion of the solvent uninfluenced by the charge distribution
of the solute\cite{Pratt:94,Tawa:94,Tawa:95,Hummer:95a,Corcelli,%
Hummer:96,Pratt:97a,Hummer:97c}.  Free energy contributions due to
non-electrostatic interactions are neglected.  We address free energy
changes due to atomic motions internal to the complexes on the basis
of gas phase determinations of the vibrational frequencies and
partition functions.  The solvation calculation treats all species as
rigid; we comment later on the consequences of this approximation and
on the possibilities for relaxing it .  The solute molecular surface
was the boundary between volume enclosed by spheres centered on all
atoms and the exterior.  The sphere radii were those determined
empirically by Stefanovich \cite{Stefanovich}, except
R$_{Fe}$=2.08~\ang for the ferric ion. The ferric ion is well buried
by the ligands of these complexes and slight variations of this latter
value were found to be unimportant. The numerical solutions of the
Poisson equation were produced with a boundary integral procedure as
sketched in references \cite{Corcelli,Pratt:97a} typically employing
approximately 16K boundary elements.  The accuracy of the numerical
solution of the dielectric model is expected to be better than a
kcal/mol for the electrostatic solvation free energy.

\section{Results and Discussion}

The pertinent energies for Fe(H$_2$O)$_6$$^{3+}$ and other species
involved in hydrolysis processes are compiled in Tables \ref{T_0} and
\ref{T_1}.  The magnitudes of the various components entering into the
gas phase hydrolysis free energy are reported in Table~\ref{T_1A},
while Table~\ref{T_1B} adds the solvation contributions to the free
energy and compares the total with experiment. Table~\ref{T_2}
summarizes geometrical information and partial atomic charges computed
from the electrostatic potential fit are presented in Table~\ref{T_3}.

\subsection{Fe(H$_2$O)$_6$$^{3+}$} 
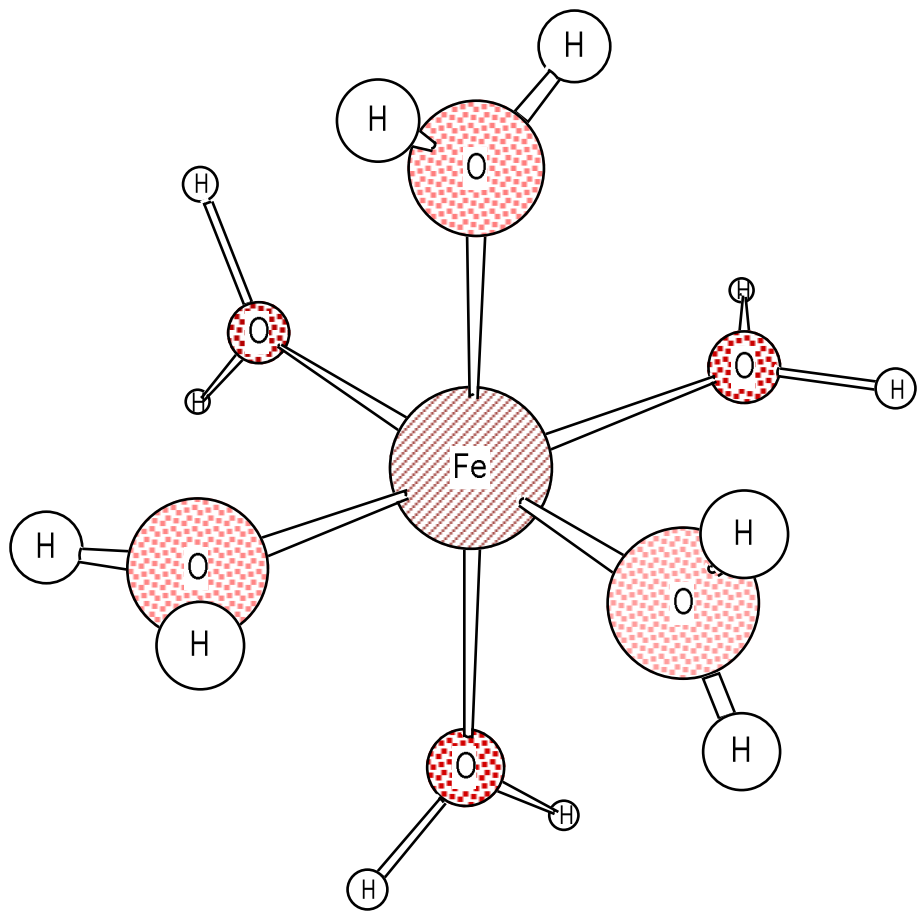
\begin{figure}

\caption{Computed structure of the Fe(H$_2$O)$_6$ complex. }

\label{fig1} \end{figure}

The structure of the
Fe(H$_2$O)$_6$$^{3+}$ complex, T$_h$ symmetry, is shown in Figure
~\ref{fig1}. The B3LYP approximation gives an Fe-O distance of
2.061~\ang. By way of comparison, with the identical basis set the
Hartree-Fock approximation yields 2.066~\ang. These distances are
similar to the recent Hartree-Fock results of {\AA}kesson, {\it et
al.\/}\cite{akesson:3+},R$_{FeO}$ = 2.062~\ang, and the
gradient-corrected DFT(BPW86) calculations of Li, {\it et
al.\/}\cite{li}, R$_{FeO}$= 2.067~\ang. These theory values cluster in
the upper end of the range of distances (1.97~\ang -- 2.05~\ang)
determined experimentally\cite{Neilson,Brunschwig,ohtaki}.  Neutron
scattering measurements report 2.01~\ang\ in concentrated electrolyte
solutions \cite{Neilson}.  The solution EXAFS result (1.98~\ang) is
close to the crystallographic determinations, 1.97~\ang --
2.00~\ang\cite{Brunschwig}.  In their recent review, Ohtaki and Radnai
discuss a number of other experiments and conclude the distance lies
in the range of 2.01 -- 2.05~\ang\ \cite{ohtaki}.

Though these theory values are compatible with the upper end of this
range, they are far from the EXAFS result (1.98~\ang) that might be
the most reliable experimental determination. While investigating this
question, we found that the B3LYP value is stable with respect to
further improvements in the basis; if the basis is augmented with an
f-function on the metal and diffuse functions on the oxygens, the
equilibrium distance decreases only slightly (R$_{Fe-O}$ = 2.053~\ang).
As noted above, the B3LYP result is in close agreement with that from
the Hartree-Fock approximation. It would not be surprising, however,
if the HF approximation overestimated the bond length.  Akesson, {\it
et al.\/}\cite{akesson:2+,akesson:more2+} have previously observed
that the HF bond lengths for a series of first row transition metal
hydrates are systematically too long. They find that correlation
effects and the influence of the second hydration shell each act to
reduce the bond length by about 0.01~\ang. Even with these corrections
the theoretical bond length is still much larger than 1.98~\ang. As an
additional data point, the local-density-approximation with the
present basis yields R$_{Fe-O}$ = 2.013~\ang, in better agreement with
the EXAFS and crystallographic determinations. However, while the LDA
generally gives reasonable geometries for transition metal
complexes\cite{ziegler,sosa}, Sosa, {\it et al.\/}\cite{sosa} find
that it tends to underestimate the lengths of dative bonds, such as
the one discussed here, and this might further support a bond length
toward the upper end of the range.

The atomic partial charges computed here for the hexaaquoferric ion are
given in Table ~\ref{T_3}.  Note that the charge assigned to the ferric
ion is substantially less than 3e, and that the magnitudes of the
charges on the oxygen and hydrogen atoms are substantially greater than
is common with effective force fields used for simulation of liquid
water.

As discussed in the appendix, a value for the absolute free energy of
the ferric ion can be obtained from a free energy of formation of the
Fe(H$_2$O)$_6$$^{3+}$ complex from an isolated Fe$^{3+}$ ($^6S$) ion
and six water molecules (Table~\ref{T_1A}), provided that no solvation
contribution is included for the atomic ion and that the actual, not
standard state, species concentrations are used.  That absolute free
energy in aqueous solution is -1020~kcal/mol. Experimental values
range from -1037~kcal/mol to -1019~kcal/mol
\cite{Rosseinsky,urban,Friedman:73,YMarcus}. We note that this
theoretical value takes no account of solvation effects due to
non-electrostatic interactions.

This agreement with experiment is encouraging.  In computing the
absolute free energy there are a number of large contributions to the
final result. This contrasts to the hydrolysis reactions where we have
arranged things to encourage cancellation of errors. It is interesting
to examine the components contributing to the free energy in the gas
phase (Table~\ref{T_1A}). Not surprisingly, the zero-point correction,
+15~kcal/mole, is significant. Under the standard conditions,
hypothetical p=1~atm ideal gas with T=298.15~K, there is the large,
unfavorable differential entropy contribution, +47~kcal/mole.  This
arises from the necessity of sequestering six water molecules in a
dilute gas. The net free energy found for the gas phase reaction is
-602~kcal/mole. Table~\ref{T_1B} addresses the solution phase aspects
of this thermochemistry. (See the Appendix also.)  About half of the
ideal entropy penalty mentioned is regained in the liquid because of
the higher concentration of water molecules.  The differential
solvation free energy adds another -391~kcal/mole of (favorable) free
energy for a net absolute free energy of hydration of -1020~kcal/mole.

This estimate of the absolute free energy of the ferric ion is close
to the value -1037~kcal/mole reported by Li, {\it et al.\/} \cite{li}
as a hydration {\it enthalpy.\/} The Li, {\it et al.,\/} value
contains some terms that would be appropriate if the hydration
enthalpy were sought and some terms that would contribute to the
hydration free energy; thus comparison of that previous value with the
present result is not straightforward. In fact, the most pragmatic
calculation of the hydration enthalpy within the dielectric continuum
model is nontrivial.  The enthalpy would be then obtained by
determining a temperature derivative of the chemical potential.  The
appropriate temperature derivative determines the enthalpy directly
or, alternatively, the solvation entropy so that the desired enthalpy
might be determined by differencing with the already known chemical
potential. That temperature derivative would generally involve also
the temperature variation of the
radii-parameters\cite{Pratt:94,Tawa:94,Tawa:95,Pratt:97a,Hummer:97c}.
The variations with thermodynamic state of the radii-parameters have
not been well studied\cite{Tawa:95}. There is good agreement, however,
in many of the components contributing to this hydration enthalpy. Li,
{\it et al.,\/} report a gas phase energy of formation of
-652.2~kcal/mole with the BPW86 gradient-corrected functional compared
to the present B3LYP result of -655.2~kcal/mole, and their solvation
free energy is -444~kcal/mole {\it vs.\/} the -441~kcal/mole found
here.  Such close agreement is encouraging, although somewhat
fortuitous. Our gas phase energy of formation includes a correction,
not considered by Li, {\it et al.,\/} of some 15~kcal/mole for the
zero-point energies. It is encouraging, however, that the estimates
disagree by only about 15~kcal/mole, or some 1.5\%. Additionally, the
definition of the molecular surface adopted by Li, {\it et al.,\/} for
the solvation calculation was substantially different from that used
here, as were the radii-parameters used.

\subsection{First and second hydrolysis reactions} 
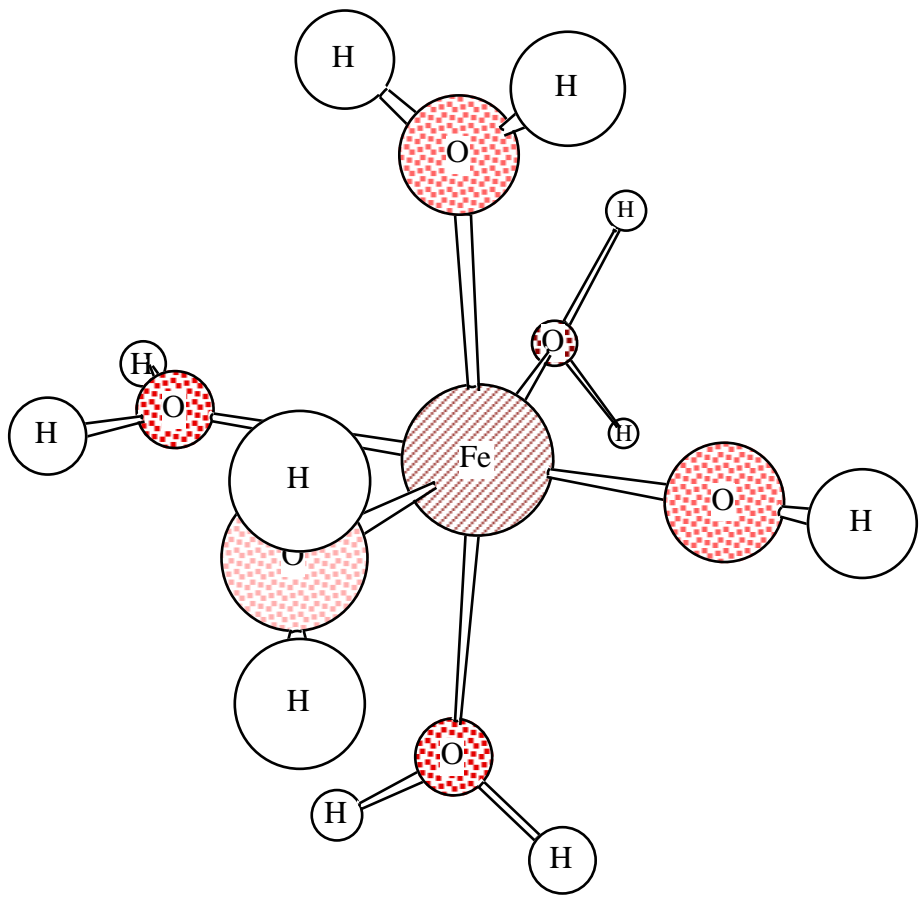
\begin{figure}

\caption{Computed structure of the Fe(H$_2$O)$_5$OH$^{2+}$ complex. }

\label{fig2} \end{figure}

We turn now to the
first deprotonation reaction Eq.~\ref{eq:hydro1}.  The structure found
for Fe(H$_2$O)$_5$OH$^{2+}$ is displayed in Figure ~\ref{fig2}. The
Fe-O (hydroxide) distance is 1.76~\ang\ and the Fe-O (water) distances
lengthen to 2.10~\ang -- 2.15~\ang. Assembling the results for the
standard free energy of this reaction we find 2~kcal/mol, in
surprisingly good agreement with the experimental value of
3~kcal/mol\cite{Flynn}.  This computed net free energy change is
composed of approximately -148~kcal/mol exothermic (favorable) change
in isolated molecule free energy and +150~kcal/mol (unfavorable) net
increase in solvation free energy.  The solvation contribution favors
the reactant side here because it presents the most highly charged
ion.  Changes in the radius assigned to the iron atom in the range
2.06~\ang$\le$R$_{Fe}\le$2.10~\ang lead to changes of $\pm$ 1~kcal/mol
in the predicted reaction free energy. This reaction was also
considered by Li, {\it et al.,\/}\cite{li} who find it to be
exothermic by 14~kcal/mole.

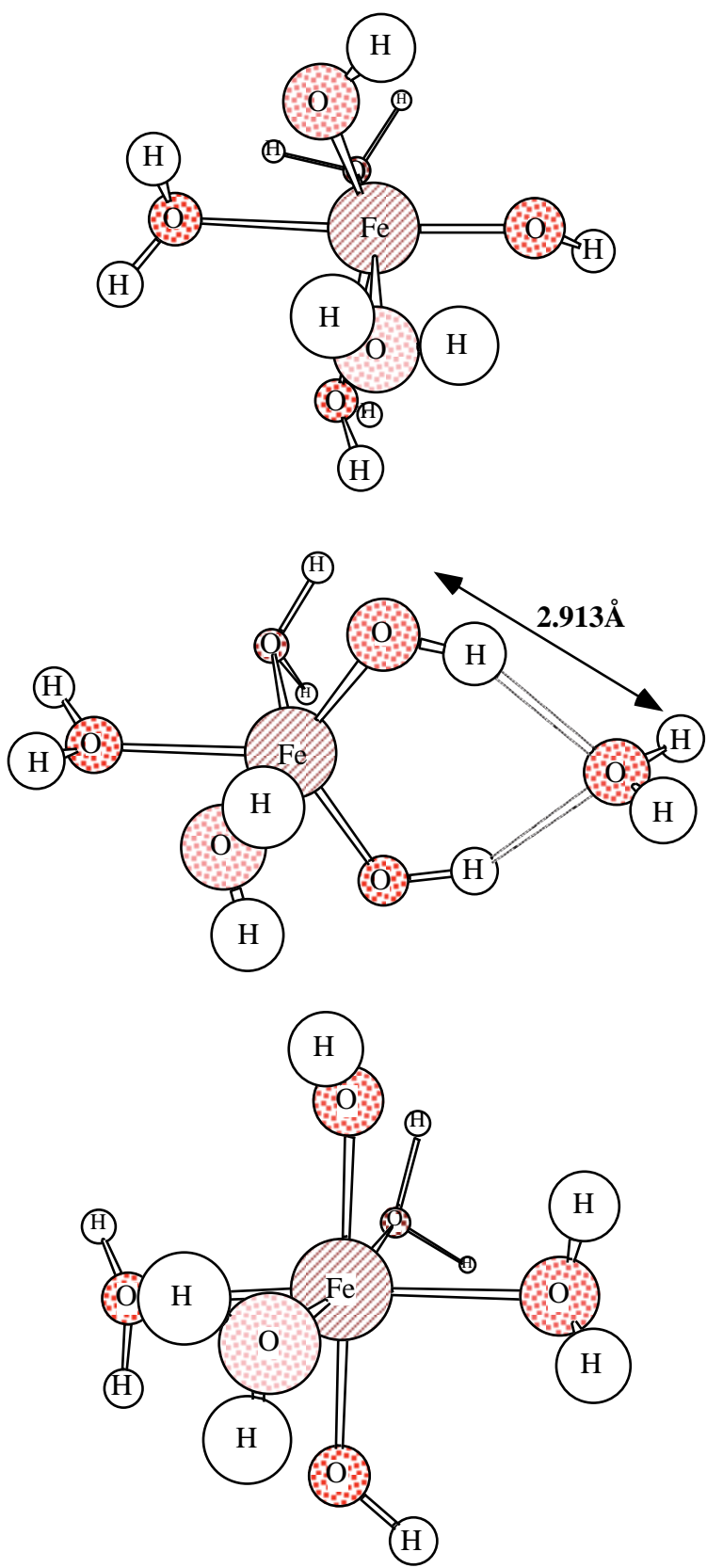
\begin{figure}

\caption{Computed structures of the Fe(H$_2$O)$_4$OH$^{2+}$ complexes;
{\it cis\/} (upper), {\it outer\/} sphere (middle), {\it trans\/} (lower), in
order from lowest to highest energy. }

\label{fig3} \end{figure}

To treat the next hydrolysis Eq.~\ref{eq:hydro2} we must consider
Fe(H$_2$O)$_4$(OH)$_2$$^+$ species.  Figure ~\ref{fig3} shows the stable
structures found.  Further lengthening of both the Fe-O(water) and
Fe-O(hydroxide) distances is noted(Table~\ref{T_2}) in the {\it cis\/}
and {\it trans\/} six-coordinate species by 0.07-0.14~\ang compared to
the Fe(H$_2$O)$_5$OH$^{2+}$.  In the gas phase, the {\it cis\/}
structure is predicted to be the lowest energy conformer, slightly
(1~kcal/mole) below the {\it trans\/} isomer. This preference is
reversed in solution, where the {\it trans\/} isomer is predicted to be
slightly more stable. We note that the most current force fields applied
to simulation of ferric ions in solution place the {\it cis\/} structure
significantly higher in energy than the {\it trans\/}\cite{Rustad:96b}.

We were surprised to discover a stable {\it outer\/} sphere complex
during our search for the stable {\it trans\/} structure.  The
structure is given in Figure ~\ref{fig3}. The distances between the
hydroxyl oxygens and the outer sphere water are typical of hydrogen
bonds. Explicit calculation of the vibrational frequencies show it to
be a true local minimum, lying less than a kcal/mol higher in energy
than the {\it cis\/} conformer.

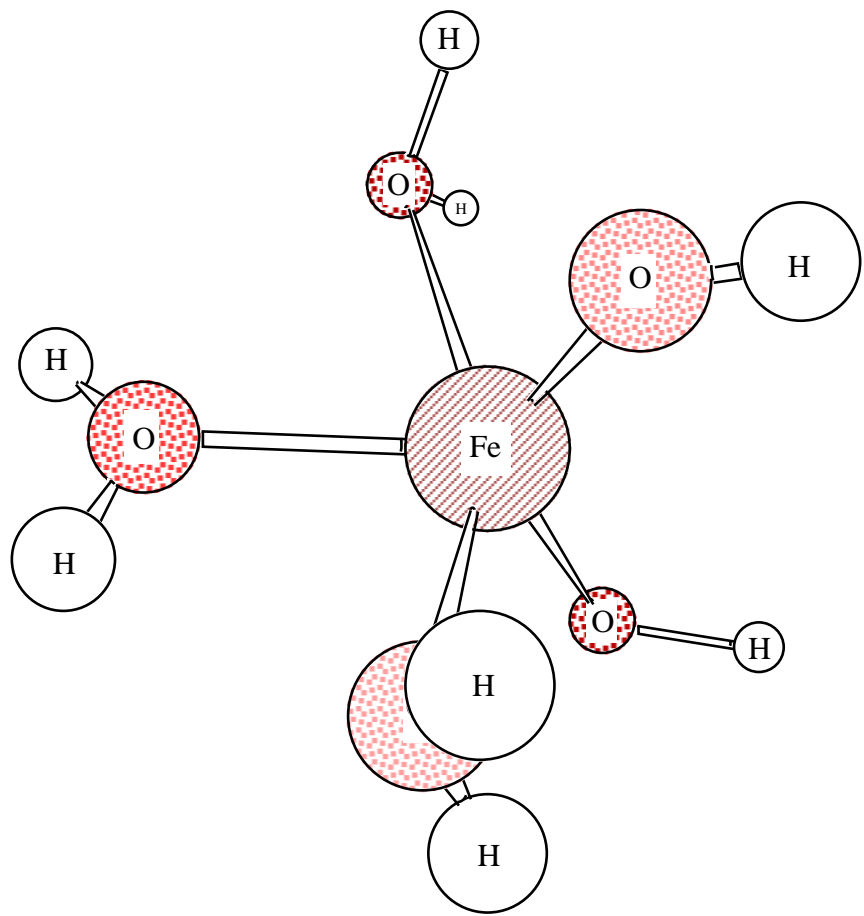
\begin{figure}

\caption{Computed structure of the penta-coordinate
Fe(H$_2$O)$_3$(OH)$_2^+$ complex.  This should be compared to the middle
structure of Figure \protect{\ref{fig3}}.}

\label{fig4} \end{figure}

The interaction of the outer sphere water with the remainder of the
ferric hydrate complex can be partially characterized by finding the
energy of that complex without the outer sphere partner.  The
structure obtained for that penta-coordinate ferric ion is shown in
Figure ~\ref{fig4}. The penta-coordinate complex can stably adopt a
conformation similar to that of Figure ~\ref{fig3} also in the absence
of the outer sphere water. In terms of the zero-point corrected
electronic energy, this {\it outer\/} sphere complex is stable with
respect to loss of the H$_2$O by about 7~kcal/mole. Consideration of
the entropic contributions to the free energy find the complex still
stable with respect to loss of H$_2$O, but the differential solvation
contributions reverse this conclusion. Thus, the dissociation of the
outer sphere complex is an essentially thermoneutral process. We
suspect such intermediates play an important role in the mechanism of
the ligand exchange with the solvent.

These three conformers lead to estimates of 16~kcal/mol, 16~kcal/mol,
and 18~kcal/mol for the reaction free energy of the second hydrolysis
for {\it trans, outer,\/} and {\it cis sphere\/} products,
respectively; the experimental value is approximately
5~kcal/mol\cite{Flynn,Meagher}.  We note that as in the case of the
first hydrolysis, the gas phase predictions lead to an exothermic
reaction; it is the differential solvation free energy that tips the
scales in favor of endothermicity.

\subsection{Role of conformational entropy} The issue of the internal
motions of the complexes, and the near degeneracy of several
conformers, raises also the issue of conformational entropy of these
species. For example, in the second hydrolysis reaction, factors such
as -RT$\ln 3$ arise if the three isomers discussed here are considered
isoenergetic.  [However, the multiplicity of isoenergetic states will
surely not be just 3 and, furthermore, entropy {\it differences\/} are
required here.]  As another example, the T$_h$ hexaaquo complex surely
has a number of low-lying structures involving the rotation of the
plane of an individual H$_2$O. More generally, this conformational
entropy would be appropriately included by computing the solvation
contribution to the chemical potential, $\mu^\ast$, of the complex
according to
\begin{eqnarray}
\mu^\ast  =   -RT\ln \sum_c x_c^{(0)} e^{-\mu^\ast(c)/RT}
\label{average} 
\end{eqnarray} 
where the sum indicated is over conformations $c$ weighted by the
normalized population, the mole fractions $ x_c^{(0)} $ of conformers
when there is no interaction between solute and solvent, and further
the summand is the Boltzmann factor of the solvation free energies,
perhaps from a physical model such as the dielectric model used here,
for each conformation. [The treatment of the thermodynamics of
flexible complexes in solution is also discussed further in the
appendix and in Reference \cite{Pratt:97b}.] The solvation
contribution to the chemical potential Eq.~\ref{average} would then be
combined with the isolated cluster partition function to obtain the
free energies of the species involved and free energy changes for the
reactions\cite{Pratt:97b}.  Finally, an entropic contribution,
including any conformational entropy, would be obtained by temperature
differentiation of the full chemical potential.  The isolated cluster
partition functions would properly include an entropy associated with
the multiplicity of isoenergetic conformational states. It is
reasonable to expect that this conformational entropy increases
progressively with hydrolysis, {\it i.e.\/} the products here are
``less ordered,'' and have higher conformational entropy than their
reactants.  It is thus significant that the predicted reaction free
energies are higher than the experiment; inclusion of conformational
entropy should lower these reaction free energies.

We note that Li, {\it et al.\/}\cite{li} argue that inclusion of a
second solvation shell substantially improves the accuracy of the
thermochemical predictions.  The present results do not seem to force
us to larger clusters.  Although it is true that proper inclusion of
more water molecules should permit a more convincing treatment,
inclusion of more distant water molecules makes the neglect of
conformational entropy less tenable. In any case,
comparison of dielectric model treatments with thermochemical results
only permits limited conclusions because of the empirical
adjustability of radii-parameters.

\section{Conclusions}

Given the balance of the large contributions that must be considered,
the observed accuracy of the computed hydrolysis reaction free
energies reactions is encouraging.  Evidently many structural
possibilities will have to be treated for a full description of ferric
ion speciation in water, eventually considering higher aggregates and
anharmonic vibrational motions of the strongly interacting water
molecules and other ligands. These issues are probably best pursued
through development of a molecular mechanics force field to screen
structures rapidly, reserving electronic structure calculations for
verification of the important structures found and refinement of the
force field. The energetic ordering found here for isomers of
Fe(H$_2$O)$_4$(OH)$_2$$^+$ is {\it cis\/} (lowest), {\it outer
sphere\/}, and {\it trans\/} (highest), but all these energies are
within about 2~kcal/mol.  Molecular mechanics force fields might be
reparameterized to account for these
results\cite{Rustad:95,Rustad:96a,Rustad:96b}. Identification of
prominent isomers should simplify and improve the modeling of the
temperature variations of thermodynamic properties. It deserves
emphasis also that substantial contributions to these reaction free
energies, and to the accumulated uncertainties, are associated with
the water partners in these reactions, {\it i.e.\/} the free energies
associated with solvation of H$_2$O and H$_3$O$^+$. A similar comment
would apply for other common oxy-acids and ligands in water, {\it
e.g.\/} carbonate, nitrate, sulfate, and phosphate. Molecular
descriptions of metal ion speciation will be incomplete without
accurate molecular characterizations of these species in aqueous
solution.

\section*{Acknowledgement}  This work was supported by the LDRD
program at Los Alamos.  LRP thanks Marshall Newton for helpful
discussions on these topics.

\section*{APPENDIX}

Here we specify with greater care some of the statistical
thermodynamic considerations relevant to solvation free energies
obtained from cluster calculations of the present variety\cite{G94}.
We note that these issues are of minor importance for the hydrolysis
reactions that are the focus of this paper. However, these
considerations become more important for the absolute free energy
reported for the aqueous ferric ion and for the free energy of
dissociation of the outer sphere complex.

\subsection*{Utilizing calculations on clusters} \noindent We first
specify more fully and explain the procedure from computing the absolute
free energy of the ferric ion given in Table~\ref{T_1B} on the basis of
cluster results obtained from density functional theory and the
dielectric continuum estimate of solvation contributions.  The result of
the development here is a formula for the chemical potential of the
ferric ion: 
\begin{eqnarray}
\mu_{{Fe^{3+}}} = RT\ln\lbrack{\rho_{{Fe^{3+}}} V\over q_M
\left\langle \left\langle e^{-\Delta U / RT}\right\rangle
\right\rangle _{0,M}} \rbrack - 6\mu_W, 
\label{answer3} 
\end{eqnarray} 
Here $\rho_{{Fe^{3+}}}$ the number density of ferric ions in the
solution, $V$ is the volume of the system, $ q_M $ is the partition
function of the isolated hexaaquoferric complex
(M=Fe(H$_2$O)$_6$$^{3+}$), $-RT\ln\langle\langle e^{-\Delta U /
RT}\rangle\rangle _{0,M} $ is the solvation free energy of the complex,
and $\mu_W$ is the chemical potential of the water.  The conclusion
drawn from this formula is that the desired absolute free energy of the
hydrated ferric ion is obtained from the chemical potential change of
the reaction Fe$^{3+}$+6H$_2$O $\rightarrow$ Fe(H$_2$O)$_6$$^{3+}$
except with the additional provisions that the chemical potentials
should be evaluated on a fully molecular basis at the water
concentration of interest and no solvation contribution should be
included for the atomic ion.  The subsequent section notes how the
isolated molecule partition functions should be used to obtain the
chemical potentials at the required concentrations.

As suggested above, the fundamental issue is the use of the cluster
calculation Fe(H$_2$O)$_6$$^{3+}$ to obtain the chemical potential,
$\mu_{Fe^{3+}}$, of the hydrated ferric ion, Fe$^{3+}$. It is clear on
physical grounds that such an approach should be advantageous when the
identification of a relevant cluster is physically obvious and when
the inner shell of the cluster requires a specialized treatment.  What
happens when the relevant cluster is not so obvious?  What about cases
when more than one cluster should be considered?  How might this
approach be justified more fully and how might the calculations be
improved?

The statistical mechanical topic underlying the considerations here is
that of association equilibrium
\cite{Stillinger:63,Pratt:76a,Pratt:76b} and is often associated with
considerations\cite{hill} of `physical clusters.' A suitable
clustering definition\cite{Stillinger:63,LaViolette:83} is required
for these discussions to be explicit and heavier statistical
mechanical formalisms can be
deployed\cite{Stillinger:63,Pratt:76a,Pratt:76b}. However, the
treatment here aims for maximal simplicity.  This argument is an
adaptation of the potential distribution theorem\cite{widom}.

In order to involve information on clusters, we express the density of
interest in terms of cluster concentrations. Thus, if the Fe$^{3+}$
ion appears only once in each cluster, {\it i.e.} if {\it mono}nuclear
clusters need be considered, then we would write 
\begin{eqnarray}
\rho_{Fe^{3+}} = \sum_M \rho_M \label{stoichiometry} 
\end{eqnarray}
where $M$ identifies a molecular cluster considered and the sum is over
all molecular clusters that can form.  A satisfactory clustering
definition\cite{Stillinger:63,LaViolette:83} insures that each ferric
ion can be assigned to only one molecular cluster, {\it e.g.\/} that a
molecular cluster with one ferric ion and six water molecules is not
counted as six clusters of a ferric ion with five water molecules. The
calculations above assumed M = Fe(H$_2$O)$_6$$^{3+}$, and that was it.
In the more general case that not all clusters are mononuclear,
Eq.~\ref{stoichiometry} would involve the obvious stoichiometric
coefficients.  The concentrations $\rho_M$ are obtained from
\begin{eqnarray} \rho_M = z_{{Fe^{3+}}}z_W^{n_M} (q_M/V) \left\langle 
\left\langle e^{-\Delta U / RT}\right\rangle \right\rangle _{0,M} .
\label{pdt} \end{eqnarray} 
$n_M$ is the number of water molecules in the cluster of type $M$,
(six in the example carried long here); $z_{{Fe^{3+}}}$ and $z_W$ are
the activity of the ferric ion and the water, respectively; {\it i.e.}
$z_\gamma = e^{\mu_\gamma/RT}$; $q_M=q_M(T)$ is a conventionally
defined canonical partition function for the cluster of type
$M$\cite{Stillinger:63,Pratt:76a,Pratt:76b,LaViolette:83}.  The
indicated average utilizes the thermal distribution of cluster and
solvent under the conditions that there is no interaction between
them. $\Delta U$ is the potential energy of interaction between the
cluster and the solvent.  In the example carried along here we do not
pay attention to any counter-ions since those issues are tangential to
the current considerations.

Eq.~\ref{pdt} is most conveniently derived by considering a grand
ensemble.  Suppose we have a definite clustering criterion: a cluster
of a ferric ion and $n_M$ water molecules is formed when exactly $n_M$
water molecules are within a specified distance $d$ of a ferric ion.
In the example we have been carrying along, water molecules with
oxygens within about $2.2$~\AA\ of a ferric ion are in chemical
interaction with the ferric ion.  It would be natural to specify
$d\le2.2$~\AA\ for clustered ferric-water(oxygen) distances.  The
average number $ <N_M>$ of such clusters is composed as
\begin{eqnarray}
\Xi(z_{{Fe^{3+}}},z_W,T,V)<N_M> &  = &   z_{{Fe^{3+}}}z_W^{n_M} \\
\nonumber & \times & \sum_{N_{{Fe^{3+}}} \ge 1, N_W \ge n_M}
N_{{Fe^{3+}}} z_{{Fe^{3+}}}^{N_{{Fe^{3+}}}-1} { N_W \choose n_M }
z_W^{N_W-n_M} Q({\bf N},V,T \vert n_M+1) \label{derive}
\end{eqnarray}
Here $\Xi(z_{{Fe^{3+}}},z_W,T,V) $ is the grand canonical partition
function; $ N_{{Fe^{3+}}}$ is the number of ferric ions in the systems
and $N_W $ is the number of water molecules; $Q({\bf N},V,T \vert
n_M+1) $ is the canonical ensemble partition function with one
specific ferric ion and $n_M$ specific water molecules constrained to
be clustered.  The binomial coefficient $ { N_W \choose n_M }$
provides the number of $n_M$-tuples of water molecules that can be
selected from $N_W$ water molecules.  Because of the particle number
factors in the summand the partition function there can also be
considered to be the partition function for $N-1$ ferric ions and
$N_W-n_M$ water molecules but with an extra $n_M+1$ objects that
constitute the cluster of interest.  A reasonable distribution of
those $n_M+1$ extraneous objects is the distribution they would have
in an ideal gas phase; the Boltzmann factor for that distribution
appears already in the integrand of the $Q({\bf N},V,T \vert n_M+1) $
and the normalizing denominator for that distribution is $n_M!
q_M(T)$. The acquired factor $n_M!$ cancels the remaining part of the
combinatorial $ { N_W \choose n_M }$. Adjusting the dummy summation
variables them leads to Eq.~\ref{pdt}.

If we were to identify an activity for an M-cluster as $z_M =
z_{{Fe^{3+}}}z_W^{n_M} $ we would obtain from Eq.~\ref{pdt} the
general statistical thermodynamic formulae of Reference
\cite{Pratt:97b}.

A virtue of the derivation of Eq.~\ref{pdt} sketched here is that the
primordial activities $z_{{Fe^{3+}}} $ and $z_W $ are clear from the
beginning.  This helps in the present circumstance where
concentrations and chemical potentials of many other species and
combinations will be of interest also.

Combining our preceding results, we obtain
\begin{eqnarray}
{\rho_{{Fe^{3+}}}\over z_{{Fe^{3+}}}} = \sum_M z_W^{n_M} (q_M/V)
\left\langle \left\langle e^{-\Delta U / RT}\right\rangle
\right\rangle_{0,M}.
\label{answer1}
\end{eqnarray} 
This is a reexpression according to clusters of a basic result known
both within the context of the potential distribution
theorem\cite{widom} and diagrammatic (mathematical) cluster
expansions.  For the latter context see Eq.~2.7 of
Reference~\cite{Pratt:76b}.  Rearranging, we obtain the desired
chemical potential 
\begin{eqnarray}
\mu_{{Fe^{3+}}} = -RT\ln\lbrack \sum_M ({z_W^{n_M}q_M\over
\rho_{{Fe^{3+}}} V}) \left\langle  \left\langle e^{-\Delta U /
RT}\right\rangle \right\rangle _{0,M} \rbrack . \label{answer2}
\end{eqnarray} 
This is the result that was sought. If higher-order clusters had been
considered with a suitable clustering definition, the final result
would have involved a more general polynomial of $z_{{Fe^{3+}}} $;
higher powers of $z_{{Fe^{3+}}} $ would appear in the sums that would
replace Eq.~\ref{stoichiometry} because of the presence of higher
powers of $z_{{Fe^{3+}}} $ in some instances of Eq.~\ref{pdt}.  A
virtue of Eq.~\ref{answer2} result is that the thermodynamic
activity of the water appears explicitly and that contribution may be
included in a variety of convenient ways, perhaps utilizing
experimental results. Furthermore, we see that there is no question
whether a particular standard state for the water is relevant or, for
example, whether only the excess part of the chemical potential of the
water is required.

Notice that if the cluster definition had been restrictive enough that
only the atomic M=Fe$^{3+}$ were present in appreciable concentration,
{\it i.e.\/} $n_M$=0 for all clusters that need be considered, then
Eq.~\ref{answer2} produces the previously known general answer for the
chemical potential of a ferric ion in solution
\cite{Pratt:97b,widom}:
\begin{eqnarray} {{\mu _{Fe^{3+}}}/ RT}=\ln [\rho _{Fe^{3+}}V/q
_{Fe^{3+}}]-\ln \left\langle {e^{-\Delta U / RT}} \right\rangle _0
\label{fun1}
\end{eqnarray} 
It would be natural to choose the electronic energy of the atomic
ferric ion as the zero of energy and to anticipate that the degeneracy
of the electronic degrees of freedom will be physically irrelevant.
Then it would be sufficient to put $V/q _{Fe^{3+}} = \Lambda
_{Fe^{3+}}^3 = (h^2 / 2 \pi m_{Fe^{3+}} RT)^{3/2} $, the cube of the
deBroglie wavelength of the ferric ion.  $ m_{Fe^{3+}} $ is the mass
of the ion, and $h$ is the Planck constant.  In any case, we define
the absolute free energy of the hydrated ferric ion as the second term
on the right $\Delta \mu_{Fe^{3+}} = -RT\ln \left\langle {e^{-\Delta U
/ RT}} \right\rangle _0 $.  In view of Eq.~\ref{answer3} we then have
\begin{eqnarray} 
\Delta\mu_{{Fe^{3+}}} = RT\ln\lbrack{q_{{Fe^{3+}}} \over q_M}\rbrack  
-RT\ln\left\langle \left\langle e^{-\Delta U /
RT}\right\rangle\right\rangle _{0,M} - 6\mu_W,
\label{answer4} 
\end{eqnarray} 
where M=Fe(H$_2$O)$_6$$^{3+}$.  This Eq.~\ref{answer4} is the formula
that was used.

A generalization of interest is the case that the solvent contains more
than one species that may complex with a specific metal ion.   For
example, suppose that ammonia may be present in addition to water, that
mixed complexes may form with ferric ion, and that these complexes have
been studied as clusters in the same way that the hexaaquoferric ion
complex was studied above.   Then the result Eq.~\ref{answer2} is
straightforwardly generalized by including the proper combinations of
activities of the additional ligands possible.

\subsection*{Standard State Modifications} \noindent Many {\it ab
initio} electronic structure packages, such as the Gaussian94 package
used here, can produce molecular (or cluster) partition functions
$q_M=q_M(T)$ and on this basis free energies for the species and
reactions considered.  It should be emphasized that these are
typically applicable to a hypothetical ideal gas at concentrations
corresponding to pressure p=1 atm, see Table~\ref{T_1A}.  Thus, for
example, these results determine the ideal chemical potential $ \mu_W
= RT\ln\lbrack
\rho_W V/q_W(T)\rbrack$. However, because of the choice of hypothetical
p=1~atm ideal gas standard state, those results use $\rho_W = p/RT $
with p=1~atm. In Gaussian94, for example, this logarithmic concentration
dependence is considered a translational entropy contribution and,
therefore, the entropic contribution to the reaction free energy of the
first reaction in Table~\ref{T_1A} is substantial and unfavorable.
Because of the dilute reaction medium associated with p=1~atm, the water
molecules written as reactants in this reaction have more freedom before
complexation than they do after. This is an entirely expected physical
effect but inappropriate in solution.  To obtain results applicable at
the concentration of liquid water we determine that pressure parameter
$p=\rho_W RT$ from the experimental density of liquid water $\rho_W $=
997.02~kg/m${^3}$.  The required value is p=1354~atm as indicated in
Table~\ref{T_1B}.  When this value is utilized in the expression for the
translational entropy contributions, the translational entropy penalty of
Table~\ref{T_1A} for the first reaction is about half recovered. 
Because the second through fifth reactions of Table~\ref{T_1A} were
written to have the same molecule numbers for reactions and products
this translational entropy contribution is very minor in those cases.

We note also that the experimental tabulation of
Reference~\cite{Friedman:73} suggested as an experimental result for the
absolute free energy of the hydrated ferric ion (-1037~kcal/mol) of
Table~\ref{T_1B} requires an adjustment of about 1.9 kcal/mol
\cite{Hummer:96,YMarcus} for similar reasons.  This adjustment is
insignificant for that property with the methods used here.

\begin{table}

\caption{Electronic energy (E/[au]), Zero-point energy
(ZPE/[kcal/mole]), and excess chemical potential
($\mu^\ast$/[kcal/mol]) with the B3LYP and dielectric continuum
solvation approximations.  Dielectric radii for all atoms were taken
from Ref.\protect{\cite{Stefanovich}} except R$_{Fe}$=2.08\AA. }

\label{T_0}

\begin{tabular}{lddd} & E & ZPE & $\mu^\ast$ \\ \hline 
H & -0.497885   & -- & -- \\ 
Fe$^{3+}$& -1261.590210  & -- & -- \\ 
H$_2$O & -76.434010 & 13.3 & -8.3 \\ 
H$_3$O$^+$ & -76.707704 & 21.5 & -96 \\ 
OH & -75.739116 & 5.2  & --  \\ 
OH$^-$ & -75.802535 & 4.5 & -108 \\ 
Fe(H$_2$O)$_6$$^{3+}$& -1721.261664& 94.3& -441  \\ 
Fe(H$_2$O)$_5$OH$^{2+}$& -1721.216989& 85.6& -203 \\ 
{\it cis\/} Fe(H$_2$O)$_4$(OH)$_2$$^+$ & -1720.987174& 79.1& -71 \\ 
{\it trans\/} Fe(H$_2$O)$_4$(OH)$_2$$^+$& -1720.983894& 78.3& -74 \\ 
{\it outer\/} Fe(H$_2$O)$_4$(OH)$_2$$^+$& -1720.985680& 79.0& -73 \\ 
Fe(H$_2$O)$_3$(OH)$_2^+$ & -1644.536430& 62.8&-69 \\
\end{tabular} \end{table}

\begin{table}

\caption{Gas-phase thermochemistries (${\Delta}E_{0}$,kcal/mole) in the
B3LYP approximation. Note that these include zero-point energy. The
ionization potential (IP) of hydrogen and electron affinity (EA) of
the hydroxyl radical are in eV.}

\label{T_1}

\begin{tabular}{ldd} & ${\Delta}E_0$ & Expt.
\protect{\cite{G2a,G2b}} \\ \hline
H$_2$O $\rightarrow$  H + OH & 115.5  & 118.0 \\
H$_2$O $\rightarrow$ H$^+$ + OH$^-$ & 387.5 & 389.3 \\
H$_2$O + H$^+$ $\rightarrow$  H$_3$O$^+$ & -163.5 & -165.1 \\ 
2H$_2$O $\rightarrow$  H$_3$O$^+$ + OH$^-$& 223.9 & 224.0 \\ \hline
IP(H)   & 13.55 & 13.6 \\
EA(OH)  & 1.76  & 1.83 \\ 
\end{tabular} \end{table}

\begin{table}
\squeezetable
\caption{Ideal gas thermochemistries (kcal/mole) at T=298.15~K and
p=1~atm, utilizing the B3LYP approximation. The first column gives the
electronic energy contribution, the second includes the zero-point
energy, the third evaluates the energy at 298.15K, the fourth gives the
enthalpy at 298.15K, and the fifth column is the Gibbs free energy.  }

\label{T_1A}

\begin{tabular}{lddddd} & ${\Delta}E_e$& ${\Delta}E{_0}$&
${\Delta}E_{298}$& ${\Delta}H_{298}^{(0)}$&${\Delta}G_{298}^{(0)}$  \\
\hline 
Fe$^{3+}$+6H$_2$O $\rightarrow$ Fe(H$_2$O)$_6$$^{3+}$ &-669.8 &-655.2
&-657.2&-660.8&-601.6\\
Fe(H$_2$O)$_6$$^{3+}$ + H$_2$O $\rightarrow$  Fe(H$_2$O)$_5$OH$^{2+}$ +
H$_3$O$^+$&-143.7 &-144.2 & -143.2&-143.2&-148.4\\
Fe(H$_2$O)$_5$OH$^{2+}$ + H$_2$O $\rightarrow$ {\it cis\/}
Fe(H$_2$O)$_4$(OH)$_2$$^+$ + H$_3$O$^+$& -27.5&-25.8&-26.6&-26.6&-25.7\\
Fe(H$_2$O)$_5$OH$^{2+}$ +H$_2$O $\rightarrow$ {\it trans\/}
Fe(H$_2$O)$_4$(OH)$_2$$^+$ + H$_3$O$^+$&-25.5 &-24.5&-25.0&-25.0&-24.9\\
Fe(H$_2$O)$_5$OH$^{2+}$+H$_2$O $\rightarrow$ {\it outer\/}
Fe(H$_2$O)$_4$(OH)$_2$$^+$+ H$_3$O$^+$& -26.6&-25.0&-25.5&-25.5&-25.9\\
{\it outer\/} Fe(H$_2$O)$_5$OH$^{2+}$ $\rightarrow$
Fe(H$_2$O)$_3$(OH)$_2$$^+$+ H$_2$O& 9.6&6.7&6.7&7.3&-1.8\\ \hline
\end{tabular} \end{table}

\begin{table}

\caption{Aqueous reaction thermochemistries (kcal/mole) at T=298.15~K
utilizing the B3LYP approximation. The first column gives the
gas-phase free energy from Table~\protect{\ref{T_1A}}, the second
reports the net free energy after adjustment for
the alternative concentration in the liquid (``p=1354~atm'' for the
water species), and the third column gives the solvation increment to
the free energy , {\it i.e.\/} that contribution due to solute-solvent
interactions.  Experimental results are given in the final column. The
solution phase results of the first row are the ``absolute free
energies of the hydrated ferric ion.'' See the appendix,
Eq.~(\protect{\ref{answer3}}) for interpretation.  }

\label{T_1B}
\squeezetable

\begin{tabular}{lddddd} & ${\Delta}G_{298}^{(0)}$&${\Delta}G_{298}$
&$\mu^\ast$& ${\Delta}G_{298}+\mu^\ast$ & Expt.\protect{\cite{G2a,G2b}}
\\ \hline
Fe$^{3+}$+6H$_2$O $\rightarrow$ Fe(H$_2$O)$_6$$^{3+}$
&-601.6&-628.9&-391.2&--1020.1&-1019,-1037 \\
Fe(H$_2$O)$_6$$^{3+}$ + H$_2$O $\rightarrow$  Fe(H$_2$O)$_5$OH$^{2+}$ +
H$_3$O$^+$& -148.4&-148.7&150.3&1.6&3\\
Fe(H$_2$O)$_5$OH$^{2+}$ + H$_2$O $\rightarrow$ {\it cis\/}
Fe(H$_2$O)$_4$(OH)$_2$$^+$ + H$_3$O$^+$&-25.7&-26.0&44.3&18.3&5  \\
Fe(H$_2$O)$_5$OH$^{2+}$ +H$_2$O $\rightarrow$ {\it trans\/}
Fe(H$_2$O)$_4$(OH)$_2$$^+$ + H$_3$O$^+$&-24.9&-25.2&41.3&16.1 \\
Fe(H$_2$O)$_5$OH$^{2+}$+H$_2$O $\rightarrow$ {\it outer\/}
Fe(H$_2$O)$_4$(OH)$_2$$^+$+ H$_3$O$^+$& -25.9&-26.1&42.3&16.3\\
{\it outer\/} Fe(H$_2$O)$_5$OH$^{2+}$ $\rightarrow$
Fe(H$_2$O)$_3$(OH)$_2$$^+$+ H$_2$O& -1.8&+2.7&-4.3&-1.6\\ \hline
\end{tabular} \end{table}

\begin{table}

\caption{Selected geometrical parameters for species in Fe(H$_2$O)$_6$
hydrolysis reactions from B3LYP calculations.}

\label{T_2}

\begin{tabular}{lddcc} & R(Fe-OH$_2$)\AA & R(Fe-OH)\AA &
$\theta$(OH-Fe-OH)  & $\theta_{tilt}$(H$_2$O)\footnote{Defined as
O-H$_1$-H$_2$-Fe dihedral angle.} \\ \hline
Fe(H$_2$O)$_6^{3+}$ & 2.060 & -- & -- & 0(6) \\ 
Fe(H$_2$O)$_5$OH$^{2+}$ & 2.103 -- 2.150 & 1.760 & -- & 0(3),30(2) \\ 
{\it cis\/} Fe(H$_2$O)$_4$(OH)$_2^{1+}$ & 2.172 -- 2.296 & 1.820 --
1.847 & 108 & 36,40,44,51 \\
{\it trans\/} Fe(H$_2$O)$_4$(OH)$_2^{1+}$ & 2.168 -- 2.194 & 1.851 & 180
& 38(2),44(2) \\
{\it outer\/} Fe(H$_2$O)$_4$(OH)$_2^{1+}$ & 2.108 -- 2.196 & 1.811 & 114
& 18,44,44 \\
Fe(H$_2$O)$_3$(OH)$_2$ & 2.102 -- 2.185 & 1.803 -- 1.807 & 128 &
22,40,40 \\
\end{tabular} \end{table}

\begin{table}

\caption{Atomic charges from ChelpG analysis of the electrostatic
potential.}

\label{T_3}

\begin{tabular}{ldddddd} & Fe(H$_2$O)$_6$ & Fe(H$_2$O)$_5$OH &
Fe(H$_2$O)$_4$(OH)$_2$ & Fe(H$_2$O)$_4$(OH)$_2$ & Fe(H$_2$O)$_4$(OH)$_2$
& Fe(H$_2$O)$_3$(OH)$_2$  \\
& & & {\it cis\/} & {\it trans\/} & {\it outer\/} & \\ \hline 
Fe  & +2.15 & +1.62 & +1.14 & +1.54 & +1.24 & +0.92 \\ 
O2  &   -1.02 & -0.88 & -0.69 & -0.85 & -0.69 & -0.63 \\ 
H8  &   +0.58 & +0.50 & +0.43 & +0.47 & +0.41 & +0.41 \\ 
H9  &   +0.58 & +0.52 & +0.42 & +0.48 & +0.42 & +0.42 \\ 
O3  &  -1.02 & -0.88 & -0.68 & -0.89 & -0.69 & -0.62 \\ 
H10 &   +0.58 & +0.50 & +0.43 & +0.49 & +0.41 & +0.41 \\ 
H11 &   +0.58 & +0.52 & +0.42 & +0.48 & +0.42 & +0.41 \\ 
O4  &   -1.02 & -0.78 & -0.62 & -0.86 & -0.66 & -0.48 \\ 
H12 &   +0.58 & +0.47 & +0.40 & +0.45 & +0.42 & +0.37 \\ 
H13 &   +0.58 & +0.47 & +0.39 & +0.48 & +0.42 & +0.37 \\ 
O5  &   -1.02 & -0.77 & -0.71 & -0.83 & -0.88 & -- \\ 
H14 &   +0.58 & +0.47 & +0.43 & +0.45 & +0.45 & -- \\ 
H15 &   +0.58 & +0.47 & +0.42 & +0.48 & +0.45 & -- \\ 
O6  &   -1.02 & -0.80 & -0.82 & -0.90 & -0.77 & -0.67 \\ 
H16 &   +0.58 & +0.47 & +0.44 & +0.45 & +0.40 & +0.39 \\ 
H17 &   +0.58 & +0.41 & --   & --   & --   & -- \\ 
O7  &   -1.02 & -0.97 & -0.82 & -0.89 & -0.75 & -0.68 \\ 
H18 &   +0.58 & +0.60 & +0.44 & +0.44 & +0.38 & +0.39 \\ 
H19 & +0.58 & --   & --   & --   & --   & -- \\ 
\end{tabular} \end{table}
\end{document}